\documentclass[aps,pra,showpacs,groupedaddress,floatfix,nofootinbib]{revtex4}

\begin{document}

\title{On nonadiabatic calculation of dipole moments}

\author{Francisco M. Fern\'andez}

\affiliation{INIFTA (UNLP, CCT La Plata-CONICET), Divisi\'on Qu\'imica
Te\'orica Blvd. 113 y 64 S/N, Sucursal 4, Casilla de Correo 16, 1900 La
Plata, Argentina}

\begin{abstract}
We show that a recent non Born--Oppenheimer calculation of dipole moments
exhibits obscure points and is not consistent with the well known
Hellmann--Feynman theorem.
\end{abstract}

\maketitle

\section{Introduction\label{sec:intro}}

Some time ago Cafiero and Adamowicz\cite{CA02a} calculated the dipole
moments of LiH and LiD without resorting to the Born--Oppenheimer (BO)
approximation. Their approach consists in the expansion of the eigenfunction
of the Coulomb Hamiltonian in a basis set of floating s--type explicitly
correlated Gaussian functions\cite{CA02a}. The dipole moments obtained in
this way are essentially identical to the experimental values\cite{WGK62}
and the authors claim that their calculations simulate experiment more
closely than any previous ones\cite{CA02a}. They also applied basically the
same approach to the calculation of molecular polarizabilities\cite
{CA02b,CA02c,CADL03}.

The purpose of this paper is to analyze that non Born--Oppenheimer
calculation of dipole moments of diatomic molecules in the light of the
Hellmann--Feynman theorem\cite{FC87}. In Section~\ref{sec:Preliminaries} we
outline some well known properties about the molecular Hamiltonian that are
necessary for the remaining sections, in Section~\ref{sec:H-F} we develop
the Hellmann--Feynman theorem for optimized variational wave functions, in
Section~\ref{sec:dipole_moments} we apply it to the calculation of dipole
moments described by Cafiero and Adamowicz\cite{CA02a} and in Section~\ref
{sec:conclusions} we draw conclusions and summarize the main results of this
paper.

\section{Preliminaries\label{sec:Preliminaries}}

In this section we outline some well known properties of the molecular
Hamiltonian that are necessary for present discussion of the nonadiabatic
calculation of dipole moments. A more detailed analysis was provided earlier
by other authors\cite{S94,SW05a}. Following Cafiero and Adamowicz\cite{CA02a}
we consider a nonrelativistic Coulomb Hamiltonian of the form\cite{S94,SW05a}
\begin{equation}
\hat{H}_{T}=\sum_{i}\frac{\hat{p}_{i}^{2}}{2m_{i}}+\sum_{i}\sum_{j>i}\frac{%
q_{i}q_{j}}{4\pi \epsilon _{0}r_{ij}}  \label{eq:HT}
\end{equation}
where $m_{i}$, $q_{i}$ and $\hat{p}_{i}$ are the mass, charge, and momentum,
respectively, of particle $i$ located at position $\mathbf{r}_{i}$, $r_{ij}=|%
\mathbf{r}_{i}-\mathbf{r}_{j}|$ is the distance between particles $i$ and $j$%
, and $\epsilon _{0}$ is the vacuum permittivity. The first step in the
treatment of such system of particles is the separation of the motion of the
center of mass which we carry out by means of the linear transformation
\begin{equation}
\mathbf{r}_{i}^{\prime }=\sum_{j}t_{ij}\mathbf{r}_{j}  \label{eq:r'=tr}
\end{equation}
so that the Hamiltonian operator (\ref{eq:HT}) becomes
\begin{equation}
\hat{H}_{T}=\sum_{j}\sum_{k}\left( \sum_{i}\frac{t_{ji}t_{ki}}{m_{i}}\right)
\hat{p}_{j}^{\prime }\hat{p}_{k}^{\prime }+\sum_{i}\sum_{j>i}\frac{q_{i}q_{j}%
}{4\pi \epsilon _{0}r_{ij}}=\hat{H}_{M}+\hat{H}_{CM}  \label{eq:HT2}
\end{equation}
where $\hat{p}_{i}^{\prime }$ is the momentum conjugate to $\mathbf{r}%
_{i}^{\prime }$. In this way we remove the motion of the center of mass
given by $\hat{H}_{CM}$ and are left with the translation--free molecular
Hamiltonian $\hat{H}_{M}$ provided we choose the transformation matrix
elements $t_{ij}$ conveniently\cite{S94,SW05a}.

Following Cafiero and Adamowicz\cite{CA02a} we can choose $\mathbf{r}%
_{1}^{\prime }=\mathbf{r}_{CM}$ and $\mathbf{r}_{j}^{\prime }=\mathbf{r}_{j}-%
\mathbf{r}_{1}$, $j>1$, where $\mathbf{r}_{CM}$ and $\mathbf{r}_{1}$ denote
the positions of the center of mass and the heaviest nucleus in the
molecule, respectively. Other authors have proposed a more symmetric
transformation: $\mathbf{r}_{1}^{\prime }=\mathbf{r}_{CM}$ and $\mathbf{r}%
_{j}^{\prime }=\mathbf{r}_{j}-\mathbf{r}_{NCM}$, where $\mathbf{r}_{NCM}$ is
the position of the center of mass of the nuclei\cite{F62}. In this case the
coordinates of the nuclei are not independent and one of them can be
expressed in terms of the others\cite{F62}. One advantage of the
transformation chosen by Cafiero and Adamowicz\cite{CA02a} is that the form
of the potential--energy function remains simple because $r_{j1}=|\mathbf{r}%
_{j}^{\prime }|$ and $r_{ij}=|\mathbf{r}_{i}^{\prime }-\mathbf{r}%
_{j}^{\prime }|$, $i,j>1$.

The inversion operator $\hat{\imath}$ changes the sign of the coordinates
and momenta of all the particles: $\hat{\imath}\mathbf{r}_{j}\hat{\imath}%
^{-1}=-\mathbf{r}_{j}$, $\hat{\imath}\hat{p}_{j}\hat{\imath}^{-1}=-\hat{p}%
_{j}$, where $\hat{\imath}^{-1}=\hat{\imath}$. It is clear that the
translation--free Hamiltonian operator is invariant under inversion $\hat{%
\imath}\hat{H}_{M}\hat{\imath}^{-1}=\hat{H}_{M}$. Therefore, if $\psi $ and $%
E$ are an eigenfunction and its eigenvalue, respectively, of $\hat{H}_{M}$
then $\hat{\imath}\hat{H}_{M}\psi =\hat{H}_{M}\hat{\imath}\psi =E\hat{\imath}%
\psi $. If the state is nondegenerate we conclude that $\hat{\imath}\psi
=\pm \psi $ because $\hat{\imath}^{2}=\hat{1}$ (the identity operator). In
particular, the ground state is nondegenerate and spherically symmetric\cite
{S94,SW05a,SW05b}.

The classical dipole moment for a neutral distribution of charges is given
by the well--known expression
\begin{equation}
\mathbf{\mu }=\sum_{j}q_{j}\mathbf{r}_{j}  \label{eq:dipole}
\end{equation}
It follows from $\hat{\imath}\mathbf{\mu }\hat{\imath}^{-1}=-\mathbf{\mu }$
and what was said above that $\left\langle \psi |\mathbf{\mu |}\psi
\right\rangle =0$ for any nondegenerate molecular state. In other words, the
outcome of the quantum--mechanical nonadiabatic calculation of the dipole
moment as the expectation value of the corresponding operator is zero for
any molecule in its ground state. In order to circumvent this difficulty,
Cafiero and Adamowicz considered the classical interaction between the
dipole moment and the field $\mathbf{\epsilon }$ thus obtaining the
semiclassical Hamiltonian
\begin{equation}
\hat{H}_{\epsilon }=\hat{H}_{M}-\epsilon \mathbf{\cdot \mu }
\label{eq:H_field}
\end{equation}
that behaves as $\hat{\imath}\hat{H}_{\epsilon }(\epsilon )\hat{\imath}^{-1}=%
\hat{H}_{\epsilon }(-\mathbf{\epsilon })$.\ If we choose $\mathbf{\epsilon }$
along the $z$ axis, $\mathbf{\epsilon =(}0,0,\epsilon )$, then $\mathbf{%
\epsilon \cdot \mu =}\epsilon \mu _{z}$. If $\psi _{\epsilon }$ and $%
E_{\epsilon }$ are an eigenfunction and its eigenvalue, respectively, of $%
\hat{H}_{\epsilon }$ we have $\hat{\imath}\hat{H}_{\epsilon }(\epsilon )\psi
_{\epsilon }=\hat{H}_{\epsilon }(-\epsilon )\hat{\imath}\psi _{\epsilon
}=E_{\epsilon }\hat{\imath}\psi _{\epsilon }$. Consequently, if the state is
nondegenerate then the energy is an even function of the field: $E_{\epsilon
}(-\epsilon )=E_{\epsilon }(\epsilon )$.

Throughout this paper we omit the fact that the Hamiltonian operator (\ref
{eq:H_field}) does not support bound states because it was not an issue in
the nonadiabatic calculation of molecular dipole moments and
polarizabilities\cite{CA02a,CA02b,CA02c,CADL03}.

\section{The Hellmann--Feynman theorem\label{sec:H-F}}

In what follows we briefly discuss the Hellmann--Feynman theorem for
variational wave functions\cite{FC87} because the nonadiabatic calculation
of dipole moments has necessarily been based on them. Consider a trial
function $\Phi $ that depends on a set of adjustable parameters $\mathbf{a}%
=\{a_{1},a_{2},\ldots ,a_{n}\}$. If we differentiate the variational energy $%
E(\mathbf{a})$ given by
\begin{equation}
E\left\langle \Phi \right| \left. \Phi \right\rangle =\left\langle \Phi
\right| \hat{H}\left| \Phi \right\rangle  \label{eq:E_var}
\end{equation}
with respect to the variational parameters and require that
\begin{equation}
\left( \frac{\partial E}{\partial a_{i}}\right) _{\mathbf{a}=\mathbf{a}%
^{opt}}=0,\,i=1,2,\ldots ,n  \label{eq:dE/da=0}
\end{equation}
then the optimal trial function satisfies
\begin{equation}
\left\langle \frac{\partial \Phi }{\partial a_{i}}\right| \hat{H}-E\left|
\Phi \right\rangle _{\mathbf{a}=\mathbf{a}^{opt}}+\left\langle \Phi \right|
\hat{H}-E\left| \frac{\partial \Phi }{\partial a_{i}}\right\rangle _{\mathbf{%
a}=\mathbf{a}^{opt}}=0,\,i=1,2,\ldots ,n  \label{eq:var_cond}
\end{equation}

Suppose that the Hamiltonian operator $\hat{H}$ depends on a parameter $%
\lambda $ (which may be a particle charge or mass, the intensity of an
applied field, etc). The optimized variational function will depend on $%
\lambda $ through the optimal parameters $\mathbf{a}^{opt}$. Therefore, if
we differentiate equation (\ref{eq:E_var}) with respect to $\lambda $
\begin{equation}
\frac{\partial E}{\partial \lambda }\left\langle \Phi \right| \left. \Phi
\right\rangle =\sum_{i=1}^{n}\left( \left\langle \frac{\partial \Phi }{%
\partial a_{i}}\right| \hat{H}-E\left| \Phi \right\rangle +\left\langle \Phi
\right| \hat{H}-E\left| \frac{\partial \Phi }{\partial a_{i}}\right\rangle
\right) \frac{\partial a_{i}}{\partial \lambda }+\left\langle \Phi \right|
\frac{\partial \hat{H}}{\partial \lambda }\left| \Phi \right\rangle
\end{equation}
and take into account the variational expression (\ref{eq:var_cond}) then we
prove that the optimized variational function satisfies the well known
Hellmann--Feynman theorem\cite{FC87}:
\begin{equation}
\frac{\partial E}{\partial \lambda }=\left\langle \frac{\partial \hat{H}}{%
\partial \lambda }\right\rangle  \label{eq:H-F}
\end{equation}
where $\left\langle \hat{A}\right\rangle =\left\langle \Phi |\hat{A}|\Phi
\right\rangle /\left\langle \Phi |\Phi \right\rangle $. Of course, any
eigenfunction of $\hat{H}$ and its corresponding eigenvalue satisfy this
relationship.

\section{Nonadiabatic calculation of dipole moments\label{sec:dipole_moments}
}

In order to calculate the dipole moment of the diatomic molecule Cafiero and
Adamowicz\cite{CA02a} fitted three points of the energy curve $E_{\epsilon
}(\epsilon )$ with a second--degree polynomial and obtained $\mu _{z}$ from
the coefficient of the linear term.
\begin{equation}
E_{\epsilon }(\epsilon )\approx e_{0}+e_{1}\epsilon +e_{2}\epsilon
^{2}+\ldots  \label{eq:epsilon_series}
\end{equation}
They resorted to polynomials of higher degree in other calculations\cite
{CA02b,CA02c,CADL03}.

According to the Hellmann--Feynman theorem (\ref{eq:H-F}) we expect that
\begin{equation}
\frac{\partial E_{\epsilon }}{\partial \epsilon }=-\left\langle \hat{\mu}%
_{z}\right\rangle _{\epsilon }  \label{eq:dE/depsilon}
\end{equation}
for the kind of variational function that Cafiero and Adamowicz\cite{CA02a}
chose to solve the Schr\"{o}dinger equation approximately. Consequently,
\begin{equation}
\left. \frac{\partial E_{\epsilon }}{\partial \epsilon }\right| _{\epsilon
=0}=e_{1}=-\left\langle \hat{\mu}_{z}\right\rangle _{0}=0  \label{eq:a1=0}
\end{equation}
unless the variational function is not fully optimized or it does not
recover the correct symmetry as $\epsilon \rightarrow 0$. This result is
consistent with the argument of Section~\ref{sec:Preliminaries} that $%
E_{\epsilon }(\epsilon )$ is an even function of $\epsilon $.

The simple arguments outlined above clearly show that a nonadiabatic
calculation, either as an expectation value or by fitting the energy with a
polynomial function of the field intensity should produce the only result of
zero dipole moment. However, Cafiero and Adamowicz\cite{CA02a} managed to
obtain the dipole moments of the diatomic molecules by means of the
following ``trick'':\cite{CA02a} ``The energy was calculated for each basis
with three electric field strengths, $\epsilon _{z}=0,-0.0016$, and $-0.0032$
a.u., and the energy curve was fitted with a second order polynomial in $%
\epsilon _{i}$. $\mu $ is then the first order coefficient of this fit.'' In
fact, the theoretical dipole moments estimated in this curious way appear to
converge smoothly towards the experimental ones as the number of basis
functions increases\cite{CA02a}. Later they explained this procedure in a
more detailed way\cite{CA02c}: ``We use the finite field approach in the
present work, i.e., we calculate the energies of the system for several
field strengths, we fit the energy as a function of the field strength with
a polynomial, and last we use the polynomial to determine the energy
derivatives with respect to the field at the zero field. As is clear from
the discussion above, the non--BO energy of a molecule at the field strength
$f$ is identical to the energy at the field $-f$, because when the direction
of the field changes the orientation of the molecule follows the field
direction. Thus, for any system in the H$_{2}$ isotopomer series the energy
is an even function of the field, and if it is approximated by a polynomial,
only even powers need to be used. This obviously results in a zero dipole
for any system if the dipole moment is determined as described above. An
alternative approach is to apply stronger fields and only use energies
calculated for positive field strengths in generating the polynomial fit. In
this case the use of both even and odd powers is appropriate. As we have
shown in our previous work on LiH\cite{CA02a}, the dipole moment derived
from our non--BO calculations with the procedure that uses only positive
fields and polynomial fits with both even and odd powers match very well the
experimental results. Thus in the present work we will show results obtained
using interpolations with even and odd--power polynomials.'' Summarizing:
Cafiero and Adamowicz\cite{CA02a,CA02c} tells us that it is possible to fit
an even function by means of a polynomial one with even and odd powers of
the variable provided one chooses sufficiently great values on the positive
side of the variable axis. It is clear that the nonzero coefficients of the
odd powers should be the product of numerical errors. Otherwise, if we apply
this argument to, for example, $\cos (f)$, then we will prove that $\sin
(0)\neq 0$. A question arises: if the outcome of the linear term in the
expansion of the energy as a function of the electric--field intensity was
due to numerical errors, why did the dipole moments estimated by Cafiero and
Adamowicz\cite{CA02a} agreed so accurately with the experimental ones?.
Notice that this same ``trick'' led to negligible dipole moments for the A$%
_{2}$ diatoms\cite{CA02c} which is consistent with the classical view that
they should have zero dipole moments: ``Applying the same approach to
homonuclear species (H$_{2}$, D$_{2}$, and T$_{2}$) should give the dipoles
identically equal to zero. In our calculations, these actually come out to $%
10^{-8}$. This small noise which entered our calculations was due in part to
the previously mentioned fact that the zero-field wave function we use is
not an eigenfunction of $\hat{J}^{2}$ as it should be. The level of noise
introduced is negligible, as $10^{-8}$ is 4 orders of magnitude smaller than
the size of the dipole moments for the heteronuclear species.''

It seems that Cafiero and Adamowicz\cite{CA02a} chose suitable trial wave
functions for the calculation of the dipole moments that did not reflect the
expected symmetry of the exact eigenfunctions at zero field strength
(otherwise the linear term of the polynomial (\ref{eq:epsilon_series}) would
have been zero). The success of their calculation was probably due to the
particular placement of the floating Gaussian functions that reflects the
ionic character of the chemist's classical picture of the molecules:\cite
{CA02a} ``Thus the centers corresponding to the hydrogen nucleus were
scattered from about 2.9 to about 3.1 bohrs. The lithium nucleus was, of
course, placed at the origin of the internal coordinate system. The
functional centers corresponding to the electrons were located primarily on
the two nuclei, with two electrons at the origin (about $0.0\pm 0.001$ bohrs
in all three directions) and two electrons near the H nucleus (about $%
3.05\pm 0.06$ bohrs) per basis set. This reflects the strong ionic character
in the lithium/hydrogen bond. The LiD non--BO wave function was optimized
starting from the converged LiH wave function.'' Therefore, it is not clear
that the variational wavefunction reflects the correct symmetry of the
system for the values of the electric--field intensity chosen for the fit.
If the variational functions were allowed to approach the symmetry of the
exact eigenfunctions of $\hat{H}_{\epsilon \rightarrow 0}=\hat{H}_{M}$ as
the field vanishes, the linear terms of the fitting polynomials would be
zero as well as the predicted dipole moments\cite{SW05b}. This may probably
be the case of the homonuclear diatoms\cite{CA02c}.

\section{Conclusions\label{sec:conclusions}}

When the field is on, the wavefunction no longer have spherical symmetry,
the expectation value of $\mathbf{\hat{\mu}}$ is nonzero and depends on the
field strength $\epsilon $. However, it is well--known that $E_{\epsilon
}(-\epsilon )=E_{\epsilon }(\epsilon )$ and the expansion of the energy
about $\epsilon =0$ should exhibit only even powers of the field strength
(as we have briefly shown in Section~\ref{sec:Preliminaries}). In other
words, the linear term should be zero in agreement with the
Hellmann--Feynman theorem (\ref{eq:a1=0}).

We have shown that if one allows the optimized variational wave function to
recover the correct spherical symmetry when the field intensity vanishes
then the linear term of the polynomial (\ref{eq:epsilon_series}) should be
zero as well as the predicted dipole moment. Apparently, the optimized
variational wave function used by Cafiero and Adamowicz\cite
{CA02a,CA02b,CA02c,CADL03} retains, when $\epsilon \rightarrow 0$, the
cylindrical symmetry it has for $\epsilon >0$. This may be due to the
particular placement of the floating gaussians in space. Another hint was
given by Cafiero et al\cite{CBA03}: ``Since the basis functions used here
are eigenfunctions of $\hat{J}_{z}$ but not $\hat{J}^{2}$, they are of
correct symmetry only for the non--zero--field calculations. In order to
maintain equal representation of all the points in the fit, though, they are
used for the zero field point as well. This should introduce only mild
contamination by excited rotational states and has not been a practical
problem in the past.'' However, that contamination may be the cause of the
occurrence of the linear term in the fit of $E_{\epsilon }(\epsilon )$. For
that reason, we believe that the apparent success of the approach proposed
by Cafiero and Adamowicz\cite{CA02a,CA02b,CA02c,CADL03} should be further
investigated.

It is clear from the following referee's comment that this problem is not
well understood:
``There is no spherical symmetry when the electric field is included, thus
there is no inconsistency in the calculations of Cafiero and Adamowicz
in this respect.'' However, from the arguments already given above it is
clear that the coefficient of the linear term should be the expectation value
of $\hat{\mu}_z$ at zero field strength that obviously corresponds to a
spherical--symmetric ground--state. The referee is one of the supporters
that $\sin(0)\neq 0$.


\begin{thebibliography}{99}
\bibitem{CA02a}  M. Cafiero and L. Adamowicz, Phys. Rev. Lett \textbf{88},
033002 (2002).

\bibitem{WGK62}  L. Wharton, L. P. Gold, and W. Klemperer, J. Chem. Phys
\textbf{37}, 2149 (1962).

\bibitem{CA02b}  M. Cafiero and L. Adamowicz, J. Chem. Phys \textbf{116},
5557 (2002).

\bibitem{CA02c}  M. Cafiero and L. Adamowicz, Phys. Rev. Lett \textbf{89},
073001 (2002).

\bibitem{CADL03}  M. Cafiero, L. Adamowicz, M. Duran, and J. M. Luis, J.
Mol. Struct. \textbf{633}, 113 (2003).

\bibitem{FC87}  F. M. Fern\'{a}ndez and E. A. Castro, Hypervirial theorems
(Springer, Berlin, Heidelberg, New York, London, Paris, Tokyo, 1987).

\bibitem{S94}  B. T. Sutcliffe, ''The Decoupling of Nuclear from Electronic
Motions in Molecules,'' in Conceptual Trends in Quantum Chemistry, edited by
E. S. Kryachko and J. L. Calais (Kluwer Academic Publishers, the
Netherlands, 1994), pp. 53.

\bibitem{SW05a}  B. T. Sutcliffe and R. G. Woolley, Phys. Chem. Chem. Phys.
\textbf{7}, 3664 (2005).

\bibitem{F62}  A Fr\"{o}man, J. Chem. Phys \textbf{36}, 1490 (1962)

\bibitem{SW05b}  B. T. Sutcliffe and R. G. Woolley, Chem. Phys. Lett.
\textbf{408}, 445 (2005).

\bibitem{CBA03}  M. Cafiero, S. Bubin, and L. Adamowicz, Phys. Chem. Chem.
Phys. \textbf{5}, 1491 (2003).
\end{thebibliography}
\end{document}